%% file: ST_RGE.tex
\documentclass[11pt,nofootinbib,floatfix,onecolumn,preprintnumbers]{revtex4-1}
\usepackage{hyperref}
\usepackage{graphicx}
\usepackage{color}
\usepackage{amsmath,amssymb}
\usepackage{soul}
\def\z2{$\mathbb{Z}_2$}

\def\gsim{\raise0.3ex\hbox{$\;>$\kern-0.75em\raise-1.1ex\hbox{$\sim\;$}}}
\def\lsim{\raise0.3ex\hbox{$\;<$\kern-0.75em\raise-1.1ex\hbox{$\sim\;$}}}

\newcommand{\sm}{{Standard Model }}

\newcommand {\ignore}[1]{}

\definecolor{avelinocolor}{rgb}{1.0,.0,.0}

\newcommand{\AddrMPP}{%
Max-Planck-Institut f\"ur Physik (Werner-Heisenberg-Institut),
F\"ohringer Ring 6, 80805 M\"unchen, Germany
}

\newcommand{\AddrMPIK}{%
Max-Planck-Institut f\"ur Kernphysik,
Saupfercheckweg 1, 69117 Heidelberg, Germany
}

\newcommand{\AddrAHEP}{AHEP Group, Instituto de F\'{i}sica Corpuscular --
  C.S.I.C./Universitat de Val\`{e}ncia, Parc Cientific de Paterna.\\
  C/Catedratico Jos\'e Beltr\'an, 2 E-46980 Paterna (Val\`{e}ncia) - Spain}

\begin{document}
%\title{Running Effects in the Singlet-Triplet Scotogenic Model}

% \preprint{IFIC/16-17}
% \preprint{MPP-2016-43}

\title{Consistency of WIMP Dark Matter as radiative neutrino mass messenger}

\author{Alexander Merle}\email{amerle@mpp.mpg.de}
\affiliation{\AddrMPP}

\author{Moritz Platscher}\email{moritz.platscher@mpi-hd.mpg.de}
\affiliation{\AddrMPIK}

\author{Nicol\'as Rojas}\email{nicolas.rojas@ific.uv.es}
\affiliation{\AddrAHEP}

\author{Jos\'e W. F. Valle}\email{jose.valle@ific.uv.es}
\affiliation{\AddrAHEP}

\author{Avelino Vicente}\email{avelino.vicente@ific.uv.es}
\affiliation{\AddrAHEP}

\begin{abstract}
The scotogenic scenario provides an attractive approach to both Dark Matter and neutrino mass generation, in which the same symmetry that stabilises Dark Matter also ensures the radiative seesaw origin of neutrino mass. However the simplest scenario may suffer from inconsistencies arising from the spontaneous breaking of the underlying \z2 symmetry. Here we show that the singlet-triplet extension of the simplest model naturally avoids this problem due to the presence of scalar triplets neutral under the \z2 which affect the evolution of the couplings in the scalar sector. The scenario offers good prospects for direct WIMP Dark Matter detection through the nuclear recoil method.
\end{abstract}

\maketitle

\section{Introduction}
\label{sec:intro}

The popularity of the Standard Model of particle physics rests upon its enormous success in explaining weak interaction phenomena~\cite{Agashe:2014kda} in terms of weak gauge boson exchange, their explicit discovery by the UA1 and UA2 experiments~\cite{Arnison:1983rp,Banner:1983jy}, and more recently the historic discovery of the Higgs boson~\cite{Aad:2012tfa,:2012gu}. However, there is a number of experimental indications showing that the \sm must be extended. Within these experimental indications we can name two: On the one hand, the neutrino oscillations, a phenomenon that is intimately connected to neutrino masses, and on the other, the existence of a large component of Dark Matter in the Universe.

The discovery that neutrino flavours change when these particles propagate, honoured with the Nobel prize in 2015, has been confirmed in a number of independent experiments and constitutes a landmark in particle physics~\cite{fukuda:1998mi,ashie:2004mr,fukuda:2002pe,ahmad:2002jz,eguchi:2002dm,Maltoni:2004ei}. By now neutrino oscillation measurements have reached the precision era with the neutrino mixing angles and their square mass differences well determined~\cite{Forero:2014bxa}.

Nevertheless, the good knowledge of the neutrino oscillation stays short of unveiling the underlying mechanism responsible for neutrino mass generation~\cite{Valle:2015pba}. The simplest operator capable of inducing Majorana neutrino mass terms is the $d=5$ Weinberg operator~\cite{weinberg:1980bf}, which can be realised in a variety of ways in terms of heavy messenger exchange in the framework of the seesaw mechanism and its low-scale variants~\cite{gell-mann:1980vs,yanagida:1979,mohapatra:1980ia,Schechter:1980gr,Schechter:1981cv,Lazarides:1980nt,Mohapatra:1986bd,Akhmedov:1995ip,Akhmedov:1995vm,Malinsky:2005bi}.

And on the other hand, the standard model of cosmology indicates that most of the Universe is made up of dark stuff. In particular Dark Matter constitutes most of the total mass in the Universe, and its existence is strongly indicated by a variety of observations on smaller scales. These suggest that galaxies and galaxy clusters in the Universe as a whole contain far more matter than what is directly observable.  Indeed, about $85\%$ of the matter of the Universe is made of a type that cannot be observed via its electromagnetic coupling~\cite{Bertone:2004pz}. This is the Dark Matter problem whose ultimate physics interpretation, just like neutrino oscillations, remains a challenge.

In an attempt to understand both phenomena, it has been suggested by Ma that the smallness of neutrino mass may have its roots on the stability of Dark Matter~\cite{Ma:2006km}, two of the major drawbacks of the \sm that require new particle physics. Indeed the scotogenic model is based on the validity of a \z2 parity symmetry which plays a double role, namely stabilising the \z2--odd Dark Matter particle on the one hand, and ensuring the radiative origin of neutrino mass on the other. This provides a very simple setting containing a Dark Matter candidate and generating a naturally suppressed neutrino mass at one-loop level. One of the ingredients of Ma's model is a new scalar doublet charged under the \z2 symmetry, similar to the inert doublet model~\cite{Deshpande:1977rw}. In addition, fermion singlets are added. In both cases, future prospects in Dark Matter direct detection experiments are challenging~\cite{Ibarra:2016dlb}.

Moreover, it has been shown that the simplest scheme suffers from a potentially severe problem, namely that loop effects~\cite{Bouchand:2012dx,Merle:2015ica} may drive the mass parameter of the inert scalar present in the model towards negative values~\cite{Merle:2015gea}. This behaviour would lead to the spontaneous breaking of the \z2 symmetry required for consistency at low energies and has thus been called the \emph{parity problem}: without the \z2 parity, the model would lose its Dark Matter candidate, and the neutrino mass would no longer come from a one-loop radiative seesaw mechanism. Here we show how this problem is naturally avoided in a simple extension of Ma's idea, the singlet-triplet scotogenic model proposed in~\cite{Hirsch:2013ola}, partly with the aim of achieving good prospects for direct Dark Matter detection in the scotogenic scenario.

The aim of the present work is to study the \z2 problem of the scotogenic models within the singlet-triplet extension. We analyse in detail how the extra ingredients of the model open up the possibility of naturally preserving the \z2 symmetry, since the inclusion of scalar triplets neutral under the \z2 will change the running of the couplings in the scalar sector. Mimicking the basic features of the supersymmetry-based WIMP scenario in a simpler and realistic way, our model can ensure an adequate production of Dark Matter in the early Universe as well as sizeable Dark Matter tree-level detection rates through the nuclear recoil method. As mentioned, apart from stabilising the lightest particle odd under the \z2 symmetry, this provides a way to realise the Weinberg operator radiatively, giving thus a way to explain both phenomena by means of simple \sm extensions potentially accessible at the LHC.

The paper is organised as follows: We start in Sec.~\ref{sec:model} by reviewing the singlet-triplet scotogenic model, where we also make a few simplifications compared to the original reference. Our main results are presented in Sec.~\ref{sec:pheno}, where we analyse the impact of the parity problem on the triplet-extended version of the scotogenic model and show how it can be naturally avoided in this extended setting. We finally conclude in Sec.~\ref{sec:conclusions}. The full set of renormalisation group equations for the singlet-triplet scotogenic model, which has been derived for the first time within this work, are listed in Appendix~A.

% {\color{red}
% \begin{itemize}
% \item{Unsolved Problems of the SM: DM and Neutrino Masses.}
% \item{The Scotogenic Models: Problems and Solutions.}
% \item{Previous works on the Scotogenic Model and its \z2 parity and
%   proposal for this work.}
% \end{itemize}
% }

\section{\label{sec:model}The Model}

Let us first review the singlet-triplet scotogenic model~\cite{Hirsch:2013ola}. The model is based on the standard gauge symmetry $\mathrm{SU(3)_c \times SU(2)_L \times U(1)_Y}$, extended by a discrete \z2 parity. In addition to the \sm leptons and quarks, both even under \z2, the model contains two additional $\mathrm{SU(2)_L}$ fermion fields: the singlet $N$ and the triplet $\Sigma$, both having vanishing hypercharge and being odd under \z2. The scalar sector of the model is extended as well, with the inclusion of the doublet $\eta$, also odd under \z2, and the real triplet $\Omega$, even under \z2. The lepton and scalar sectors of the model, as well as the charge assignment under $\mathrm{SU(2)_L}$, $\mathrm{U(1)_Y}$ and \z2, are shown in Table~\ref{tab:MatterModel}.

In this paper we will use the standard $2 \times 2$ matrix notation for the $\mathrm{SU(2)_L}$ triplets, which can (for vanishing hypercharge) be decomposed as
\begin{equation}
\Sigma = \left( \begin{array}{cc}
\frac{\Sigma^0}{\sqrt{2}} & \Sigma^+ \\
\Sigma^- & -\frac{\Sigma^0}{\sqrt{2}}
\end{array} \right) \, , \quad \Omega = \left( \begin{array}{cc}
\frac{\Omega^0}{\sqrt{2}} & \Omega^+ \\
\Omega^- & -\frac{\Omega^0}{\sqrt{2}}
\end{array} \right) \, . \label{eq:triplets}
\end{equation}

\begin{table}[!t]
\centering
\begin{tabular}{|c||c|c|c||c|c||c|c|}
\hline
        & \multicolumn{3}{|c||}{Standard Model} &  \multicolumn{2}{|c||}{Fermions}  & \multicolumn{2}{|c|}{Scalars}  \\
        \cline{2-8}
        &  $L$  &  $e$  & $\phi$  & $\Sigma$ &  $N$   & $\eta$ & $\Omega$ \\
\hline     
Generations & 3 & 3 & 1 & 1 & 1 & 1 & 1 \\ \hline                      
$\mathrm{SU(2)_L}$ &  2    &  1    &    2    &     3    &  1   &    2   &    3     \\
$\mathrm{U(1)_Y}$     & -1/2    &  -1    &    1/2    &     0    &  0   &    1/2   &    0  \\
\z2   &  $+$  &  $+$  &   $+$  &  $-$   & $-$  &  $-$  &  $+$    \\
\hline
\end{tabular}
\caption{\label{tab:MatterModel}Matter content and quantum numbers of the singlet-triplet scotogenic model.}
\end{table}

The most general $\mathrm{SU(3)_c \times SU(2)_L \times U(1)_Y}$, Lorentz and \z2 invariant Yukawa Lagrangian is given by
\begin{equation}
- \mathcal{L}_Y = Y_e^{\alpha \beta}\,\overline{L_{\alpha}} \, \phi \, e_{\beta} + Y_{N}^\alpha \, \overline{L_{\alpha}} \, \tilde{\eta} \, N + Y_{\Sigma}^\alpha \, \overline{L_{\alpha}} \, \tilde{\eta} \, \Sigma + Y_{\Omega} \, \overline{\Sigma} \, \Omega \, N + h.c. \label{eq:yukawa}
\end{equation}
Here, gauge contractions are omitted for the sake of compactness, flavour indices $\alpha,\beta=1,2,3$ are indicated explicitly, and we denote $\tilde{\eta} = i\sigma_2 \eta^{*}$, as usual. The $\Sigma$ and $N$ fermions are allowed to have Majorana mass terms,
\begin{equation}
- \mathcal{L}_M = 
\frac{1}{2} \, M_\Sigma \, \overline{\Sigma^{c}} \Sigma
+ \frac{1}{2} \, M_N \, \overline{N^{c}} N 
+ h.c. \label{eq:mass}
\end{equation}
Finally, the scalar potential of the model is given by
\begin{eqnarray}
\mathcal V &=& -m_{\phi}^2 \phi^\dagger \phi + m_{\eta}^2 \eta^\dagger \eta + \frac{\lambda_1}{2} \left( \phi^\dagger \phi \right)^2 + \frac{\lambda_2}{2} \left( \eta^\dagger \eta \right)^2 + \lambda_3 \left( \phi^\dagger \phi \right)\left( \eta^\dagger \eta \right) \nonumber \\ 
  &+& \lambda_4 \left( \phi^\dagger \eta \right)\left( \eta^\dagger \phi \right) + \frac{\lambda_5}{2} \left[ \left(\phi^\dagger \eta \right)^2 + \text{h.c.} \right] - \frac{m_\Omega^2}{2} \, \Omega^\dagger \Omega  \nonumber \\
  &+& \frac{\lambda^{\Omega}_1}{2} \left( \phi^\dagger \phi \right) \left( \Omega^\dagger \Omega\right) + \frac{\lambda^{\Omega}_2}{4} \, (\Omega^\dagger \Omega )^2 + \frac{\lambda^{\eta}}{2} \left( \eta^\dagger \eta \right) \left( \Omega^\dagger \Omega\right) \nonumber \\
&+& \mu_1 \, \phi^\dagger \, \Omega \, \phi + \mu_2 \, \eta^\dagger \, \Omega \, \eta \, . \label{eq:scpot}
\end{eqnarray}
Before moving on to discussing theoretical constraints on the scalar potential,  we note that our notation for the Lagrangian in Eqs.~\eqref{eq:yukawa}, \eqref{eq:mass}, and~\eqref{eq:scpot} differs slightly from the one in Ref.~\cite{Hirsch:2013ola} in two ways: (i) the scalar potential has been rewritten, removing some redundant terms and renaming the remaining ones, and (ii) the normalisation of some couplings and mass terms is different. Moreover, the triplets $\Sigma$ and $\Omega$ also have a different normalisation, as it is shown in the Eq.~\eqref{eq:triplets}.

\subsection{Theoretical constraints}
\label{subsec:theoconstraints}

The couplings in the scalar potential in Eq.~\eqref{eq:scpot} are subject to a number of constraints originating solely from theoretical considerations to be outlined in this subsection. First, we should ensure that the potential is bounded from below, as otherwise there is no stable minimum around which a perturbative expansion is feasible. The second constraint originates from this expansion being perturbatively valid, i.e.\ that the scalar quartic couplings in Eq.~\eqref{eq:scpot} are $\lesssim \mathcal{O}(1)$.

In the \sm only a single condition is necessary and sufficient for the potential to be bounded from below, namely that the Higgs quartic coupling be positive, $\lambda > 0$. Adding a second Higgs doublet complicates the situation: simple algebraic relations that ensure the boundedness cannot be found unless further symmetry assumptions are made, e.g.\ an additional $\mathbb{Z}_2$ parity under which the two doublets have different quantum numbers, cf.\ Refs.~\cite{Maniatis:2006fs, Branco:2011iw}.

Given that, in the present model, we have two scalar doublets and a triplet, finding analytic criteria for the boundedness from below of the potential is rather involved. As was noted before, the most general scalar potential allowed by the symmetries of the model contains redundant terms that have been removed in Eq.~\eqref{eq:scpot} by appropriate redefinitions of the couplings $\lambda_1^\Omega,\lambda_2^\Omega,\lambda^\eta$. Consequently, the scalar potential is a function of the real and \emph{positive} field bilinears
\begin{equation}
  h_1^2\equiv \phi^\dag \phi,\quad h_2^2\equiv \eta^\dag \eta,\quad h_3^2\equiv \mathrm{tr}\left[\Omega^\dag \Omega\right].
\end{equation}
In addition, the mixed bilinear $h_{12}^2 =\eta^\dag \phi$ can be parametrised as $h_{12}^2 = |h_1| |h_2| \rho e^{i \phi}$, with $|\rho| < 1$ by virtue of the Cauchy-Schwarz inequality, $0 \le \left|\eta^\dag \phi\right| \le |\eta| |\phi| $.

Thus, one can write the condition of boundedness from below as 
\begin{equation}
  \mathcal{V}_4 = \left(h_1^1, h_2^2, h_3^2\right) V_4 \left(\begin{array}{c} h_1^2\\ h_2^2\\ h_3^2 \end{array}\right) \ge 0,
\end{equation}
in which the matrix of quartic couplings $V_4$ is given by
\begin{equation}
  V_4 = \frac{1}{2}\left(
	\begin{array}{ccc}
	 \lambda_1 & \lambda_3 + \rho^2 \left(\lambda_4 - |\lambda_5|\right) & \frac{1}{2} \lambda_1^\Omega \\
	 \lambda_3 + \rho^2 \left(\lambda_4 - |\lambda_5|\right) & \lambda_2 & \frac{1}{2} \lambda^\eta \\
	 \frac{1}{2} \lambda_1^\Omega & \frac{1}{2} \lambda^\eta & \frac{1}{2} \lambda_2^\Omega\\
	\end{array}
	\right).
\end{equation}
In this expression, the phases $\phi$ and $\mathrm{arg}(\lambda_5)$ have been chosen such that the term proportional to $\lambda_5$ is minimal.\footnote{This term is given by $\frac{1}{2} \left( \lambda_5 h_{12}^4 + {\lambda_5}^* {h_{12}^4}^* \right) = h_1^2 h_2^2 \rho^2 |\lambda_5| \cos(2 \phi + \mathrm{arg}(\lambda_5)) \ge - h_1^2 h_2^2 \rho^2|\lambda_5|$.}

The condition $x^T V_4 x \ge 0$ for $x_i=h_i^2\ge0$ is known as \emph{co-positivity} of the matrix $V_4$, which has been well described in Ref.~\cite{Kannike:2012pe}. Using the approach outlined in this reference, necessary and sufficient conditions for the scalar potential~\eqref{eq:scpot} to be bounded from below can be obtained. In the case where $\lambda_4 + |\lambda_5| \ge 0$, we can set $\rho^2 = 0$ --~the minimum of the potential as a function of $\rho^2$~-- and in the opposite case, where $\lambda_4 + |\lambda_5| < 0$, we may fix $\rho^2=1$. This yields the conditions:
\begin{subequations}\label{eq:boundedfrombelow}
\begin{gather}
  \lambda_1 \ge 0,\qquad  \lambda_2 \ge 0,\qquad  \lambda_2^\Omega \ge 0, \\
  \lambda_3 + \sqrt{\lambda_1 \lambda_2} \ge 0,\qquad \lambda_{345} + \sqrt{\lambda_1 \lambda_2} \ge 0, \label{eq:la3bounded}\\
  \lambda_1^\Omega + \sqrt{2\lambda_1 \lambda_2^\Omega} \ge 0,\qquad \lambda^\eta + \sqrt{2\lambda_2 \lambda_2^\Omega} \ge 0,
\end{gather}
where we have used $\lambda_{345} \equiv \lambda_3 + \lambda_4 - |\lambda_5|$. Finally, we have one more condition:
\begin{equation}
  \sqrt{2\lambda_1\lambda_2\lambda_2^\Omega} + \lambda_3\sqrt{2\lambda_2^\Omega} + \lambda_1^\Omega \sqrt{\lambda_2} + 
      \lambda^\eta \sqrt{\lambda_1} + \sqrt{\left(\lambda_3+\sqrt{\lambda_1\lambda_2}\right)\left( \lambda_1^\Omega + \sqrt{2\lambda_1 \lambda_2^\Omega}\right)\left(\lambda^\eta+\sqrt{2\lambda_2\lambda_2^\Omega}\right)} \ge 0,
\end{equation}
\end{subequations}
where -- as in Eq.~\eqref{eq:la3bounded} -- we should replace $\lambda_3 \mapsto \lambda_{345}$ in case that $\lambda_4 + |\lambda_5| < 0$.

Finally, note that considering field configurations of components of $\phi,\ \eta, \textrm{ or } \Omega$ will yield equivalent or redundant expressions to Eqs.~\eqref{eq:boundedfrombelow}, because the $h_{1,2,3}^2$ are all $\mathrm{SU(2)_L}$ invariant, as pointed out in Ref.~\cite{Kannike:2012pe}.\footnote{Such an approach could be useful in a case where more ``unphysical'' parameters such as $\rho$ appear in the matrix $V_4$, as e.g.\ a parameter that describes the interdependence of the (in this setting redundant) operators $\mathrm{tr}\left[\left(\Omega^\dag \Omega\right)^2\right]$ and $\mathrm{tr}\left[\left(\Omega^\dag \Omega\right)\right]^2$, cf.~Ref.~\cite{Chakrabortty:2013mha}. However, in the present situation such interdependences are absent.}

\subsection{Symmetry breaking}
\label{subsec:scalar}

We will assume the following symmetry breaking pattern:
\begin{equation}
\langle \phi^0 \rangle = \frac{v_\phi}{\sqrt{2}} \, , \quad \langle \Omega^0 \rangle = v_\Omega \, , \quad \langle \eta^0 \rangle = 0 \, , \label{eq:vevs}
\end{equation}
with $v_\phi, v_\Omega \ne 0$. These vacuum expectation values (VEVs) are restricted by the tadpole equations
\begin{eqnarray}
t_\phi &=& -m_\phi^2 \, v_\phi + \frac{1}{2} \lambda_1 v_\phi^3 + \frac{1}{2} \lambda_1^\Omega v_\phi v_\Omega^2 - \frac{1}{\sqrt{2}} \, v_\phi v_\Omega \, \mu_1 = 0 \, , \label{eq:tad1} \\
t_\Omega &=& -m_\Omega^2 \, v_\Omega + \lambda_2^\Omega v_\Omega^3 + \frac{1}{2} \lambda_1^\Omega v_\phi^2 v_\Omega - \frac{1}{2 \sqrt{2}} v_\phi^2 \, \mu_1 = 0 \, , \label{eq:tad2}
\end{eqnarray}
obtained from the scalar potential in Eq.~\eqref{eq:scpot}, i.e. $t_i \equiv \frac{\partial \mathcal V}{\partial v_i}$ is the tadpole of $v_i$. Given the non-trivial $\phi$ and $\Omega$ charges under $\mathrm{SU(2)_L}$, the $v_\phi$ and $v_\Omega$ VEVs contribute to the $W$ and $Z$ masses,
\begin{eqnarray}
m_W^2 &=& \frac{1}{4} \, g^2 \left( v_\phi^2 + 4 \, v_\Omega^2 \right) \, , \\
m_Z^2 &=& \frac{1}{4} \left(g^2 + g'^2 \right) v_\phi^2 \, .
\end{eqnarray}
We estimate that $v_\Omega$ cannot be larger than $4.5$~GeV@$3\sigma$~\cite{Agashe:2014kda} in order to be compatible with electroweak precision tests, in particular those coming from the measurement of the $\rho$ parameter.

Let us now comment on the scalar spectrum of the model. In the basis $\text{Re} \left( \phi^0\, ,\, \Omega^0 \right)$ the mass matrix for the \z2-even and CP--even neutral scalars is given by
\begin{eqnarray}
\mathcal{M}_S^2 &=& \left(\begin{array}{cc}
-m_\phi^2 + \frac{3}{2} \lambda_1 v_\phi^2 + \frac{1}{2} \lambda_1^\Omega v_\Omega^2 - \frac{1}{\sqrt{2}} v_\Omega \, \mu_1
& \lambda^\Omega_1 v_\phi v_\Omega - \frac{1}{\sqrt{2}} v_\phi \, \mu_1 \\
\lambda^\Omega_1 v_\phi v_\Omega - \frac{1}{\sqrt{2}} v_\phi \, \mu_1
& -m_\Omega^2 + \frac{1}{2} \lambda^\Omega_1 v_\phi^2 + 3 \lambda^\Omega_2 v_\Omega^2
\end{array}\right) \, .
\end{eqnarray}
The lightest of the $S$ mass eigenstates, $S_1 \equiv h$, is identified with the $125$ GeV state recently discovered at the LHC~\cite{Aad:2012tfa,:2012gu}. Regarding the \z2-even charged scalars, their mass matrix in the basis $\left( \phi^\pm\, ,\, \Omega^\pm \right)$ can be written as
\begin{eqnarray}
\mathcal{M}_{H^\pm}^2 &=& \left(\begin{array}{cc}
-m_\phi^2 + \frac{1}{2} \lambda_1 v_\phi^2 + \frac{1}{2} \lambda^\Omega_1 v_\Omega^2 + \frac{1}{\sqrt{2}} v_\Omega \, \mu_1 + \frac{1}{4} g^2 v_\phi^2 \xi_{W^\pm} 
& \frac{1}{\sqrt{2}} v_\phi \, \mu_1 - \frac{1}{2} g^2 v_\phi v_\Omega \xi_{W^\pm} \\
\frac{1}{\sqrt{2}} v_\phi \, \mu_1 - \frac{1}{2} g^2 v_\phi v_\Omega \xi_{W^\pm}
& -m_\Omega^2 + \frac{1}{2} \lambda^\Omega_1 v_\phi^2 + \lambda^\Omega_2 v_\Omega^2 + g^2 v_\Omega^2 \xi_{W^\pm}
\end{array}\right) \, . \nonumber \\
\end{eqnarray}
Finally, we comment on the \z2-odd scalars $\eta^{0,\pm}$ states. First, we decompose the neutral $\eta^0$ field in terms of its CP-even and CP-odd components as
\begin{equation}
\eta^0 = \frac{1}{\sqrt{2}} \left( \eta^R + i \, \eta^I\right) \, .
\label{eq:defetaRI}
\end{equation}
Due to the conservation of the \z2 symmetry, the $\eta^{R,I,\pm}$ fields do not mix with the rest of scalars. Their masses are given by
\begin{eqnarray}
m_{\eta^R}^2 &=& m_{\eta}^2 + \frac{1}{2}\left(\lambda_3 + \lambda_4 + \lambda_5 \right) v_\phi^2 + \frac{1}{2}\lambda^\eta v_\Omega^2 - \frac{1}{\sqrt{2}} \, v_\Omega \, \mu_2 \, ,\label{eq:etaRmass} \\
m_{\eta^I}^2 &=& m_{\eta}^2 + \frac{1}{2}\left(\lambda_3 + \lambda_4 - \lambda_5 \right) v_\phi^2 + \frac{1}{2}\lambda^\eta v_\Omega^2 - \frac{1}{\sqrt{2}} \, v_\Omega \, \mu_2 \, ,\label{eq:etaImass}\\
m_{\eta^{\pm}}^2 &=& m_{\eta}^2 + \frac{1}{2}\lambda_3 v_\phi^2 + \frac{1}{2}\lambda^\eta v_\Omega^2 + \frac{1}{\sqrt{2}} \, v_\Omega \, \mu_2 \, . \label{eq:etaChargedmass}
\end{eqnarray}
We note that the mass difference $m_{\eta^R}^2-m_{\eta^I}^2 = \lambda_5 \, v_\phi^2$ is controlled by the $\lambda_5$ coupling and vanishes for $\lambda_5 = 0$. In this limit lepton number is recovered making the neutrinos massless, as shown below.

Finally, we emphasise that the vacuum in Eq.~\eqref{eq:vevs} preserves the \z2 scotogenic parity. This implies the existence of a stable neutral particle which can play the role of the Dark Matter of the Universe.

\subsection{Neutrino masses}
\label{subsec:numass}

The \z2-odd fields $\Sigma^0$ and $N$ get mixed by the Yukawa coupling $Y_\Omega$ and the triplet VEV, $v_\Omega$. In the basis $\left( \Sigma^0, N \right)$, their $2 \times 2$ Majorana mass matrix takes the form
\begin{equation}
\mathcal{M}_\chi = \left(\begin{array}{cc} M_\Sigma & Y_\Omega v_\Omega \\ 
Y_\Omega v_\Omega & M_N \end{array}\right) \, .
\end{equation}
The mass eigenstates $\chi_{1,2}$ are obtained after rotating to the mass basis via the $2\times2$ orthogonal matrix $V(\alpha)$,
\begin{equation}
\left(\begin{array}{c}\chi_1\\ \chi_2\end{array}\right) = \left( \begin{array}{cc}
\cos \alpha & \sin \alpha \\
-\sin \alpha & \cos \alpha
\end{array} \right) \, \left(\begin{array}{c} \Sigma^0\\ N\end{array}\right) = V(\alpha)\left(\begin{array}{c} \Sigma^0\\ N\end{array}\right),
\end{equation}
such that
\begin{equation}
\tan(2\alpha) = \frac{2 \, Y_\Omega v_\Omega}{M_\Sigma - M_N} \, .
\end{equation}

\begin{figure}[t]
\centering
\includegraphics[scale=0.45]{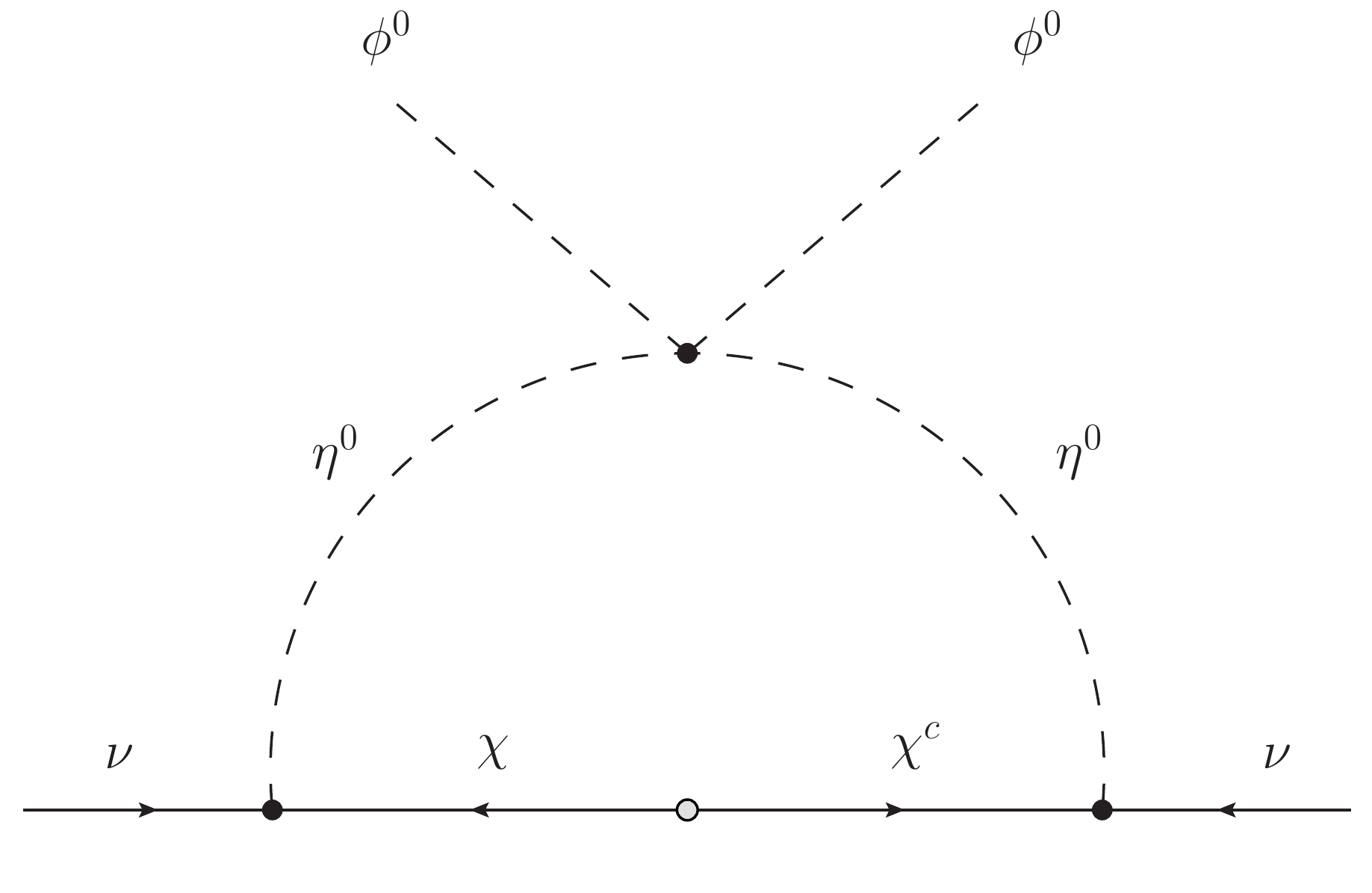}
\caption{\label{fig:numass}1-loop neutrino mass in the singlet-triplet scotogenic model. Here $\eta^0 \equiv \left( \eta^R, \eta^I \right)$ and $\chi \equiv \left( \chi_1, \chi_2 \right)$.}
\end{figure}

The singlet-triplet scotogenic model generates neutrino masses at the 1-loop level, as shown in FIG.~\ref{fig:numass}. We emphasise that this figure actually includes four 1-loop diagrams, since $\eta^0 \equiv \left( \eta^R, \eta^I \right)$ and $\chi \equiv \left( \chi_1, \chi_2 \right)$. The resulting neutrino mass matrix can be written as\footnote{We include a factor of $1/2$ that was missing in~\cite{Hirsch:2013ola}.}
\begin{eqnarray}
(\mathcal{M}_\nu)_{\alpha\beta}&=&\sum_{\sigma=1}^2\left(\frac{ih_{\alpha \sigma}}{\sqrt{2}}\right)\left(\frac{-ih_{\beta \sigma}}{\sqrt{2}}\right)\left[I(M_{\chi_\sigma}^2,m_{\eta^R}^2)-I(M_{\chi_\sigma}^2,m_{\eta^I}^2)\right] \nonumber \\
&=&\sum_{\sigma=1}^2 \frac{h_{\alpha \sigma} \, h_{\beta \sigma} \, M_{\chi_\sigma}}{2 \, (4\pi)^2} \left[\frac{m_{\eta^R}^2\ln\left(\frac{M_{\chi_\sigma}^2}{m_{\eta^R}^2}\right)}{M_{\chi_\sigma}^2-m_{\eta^R}^2} -\frac{m_{\eta^I}^2\ln\left(\frac{M_{\chi_\sigma}^2}{m_{\eta^I}^2}\right)}{M_{\chi_\sigma}^2-m_{\eta^I}^2}\right] \, , \label{eq:mnu}
\end{eqnarray}
where $h$ is a $3 \times 2$ matrix defined as
\begin{equation}
h=\left(\begin{array}{cc} 
\frac{Y_\Sigma^1}{\sqrt{2}} & Y_N^1 \\
\frac{Y_\Sigma^2}{\sqrt{2}} & Y_N^2 \\
\frac{Y_\Sigma^3}{\sqrt{2}} & Y_N^3
\end{array}\right) \cdot V^T(\alpha) \, ,
\label{eq:h-matrix}
\end{equation}
and $I(m_1^2,m_2^2)$ is a Passarino-Veltman function evaluated in the limit of zero external momentum. We note that $m_{\eta^R}^2 = m_{\eta^I}^2$ leads to vanishing neutrino masses due to an exact cancellation between the $\eta^R$ and $\eta^I$ loops. This was indeed expected, since the special limit $m_{\eta^R}^2 = m_{\eta^I}^2$ is equivalent to $\lambda_5 = 0$, in which case one can define a conserved lepton number. As a consequence of this, the choice $\lambda_5 \ll 1$ becomes natural in the sense of 't Hooft~\cite{'tHooft:1979bh}, since the limit $\lambda_5 \to 0$ enhances the symmetry of the model.

\section{\label{sec:pheno}Numerical analysis}

We now discuss the running of the model parameters numerically, where we closely follow the approach of Ref.~\cite{Merle:2015gea}. The reader is referred to this reference concerning the technical details.

\begin{figure}[!h]
  \centering
%  \begin{subfigure}{.45\textwidth}
%    \centering
    \includegraphics[width=.45\textwidth]{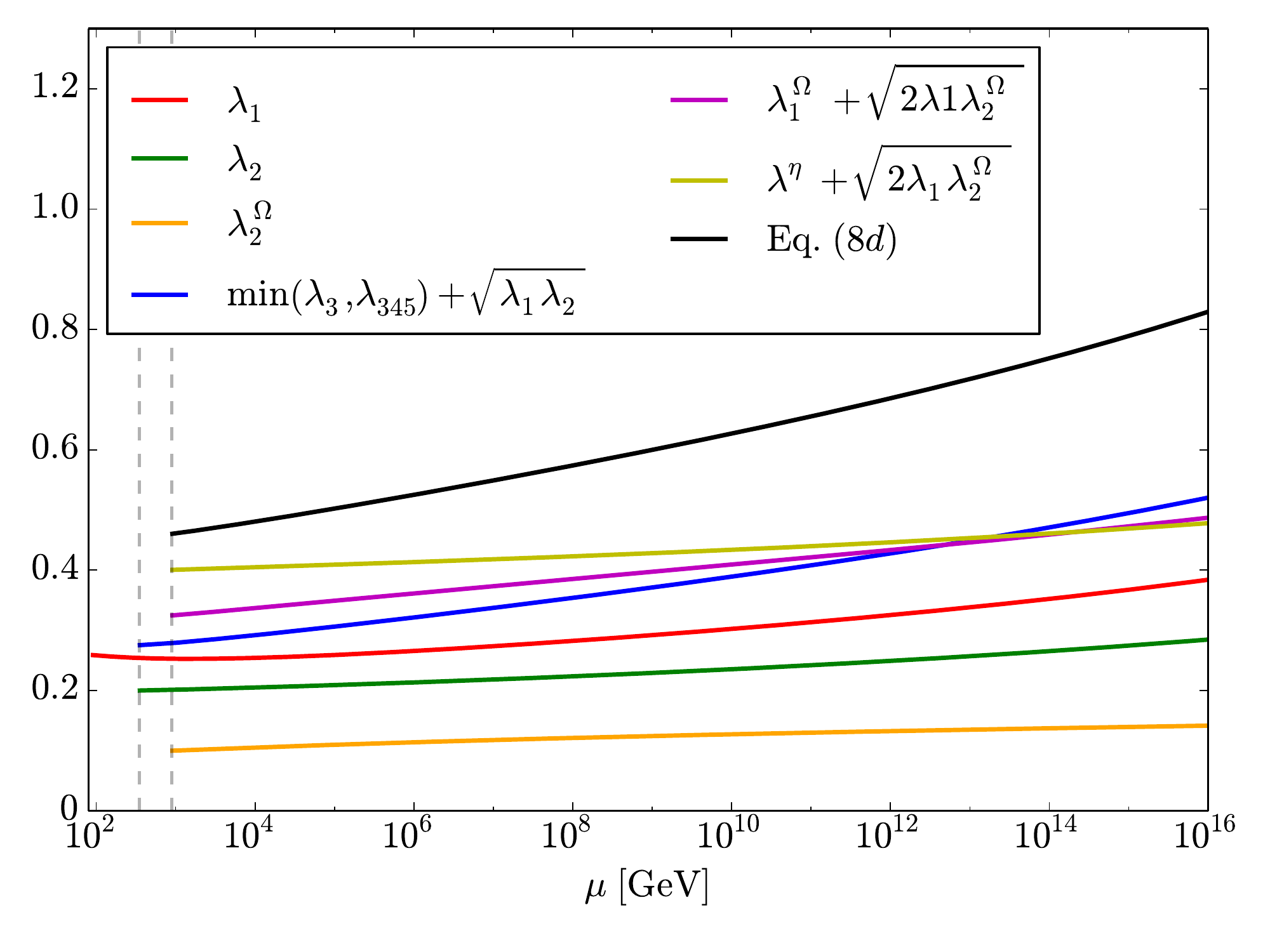}
%  \end{subfigure}
%   
 \hfill
%   
%  \begin{subfigure}{.45\textwidth}
%    \centering
    \includegraphics[width=.45\textwidth]{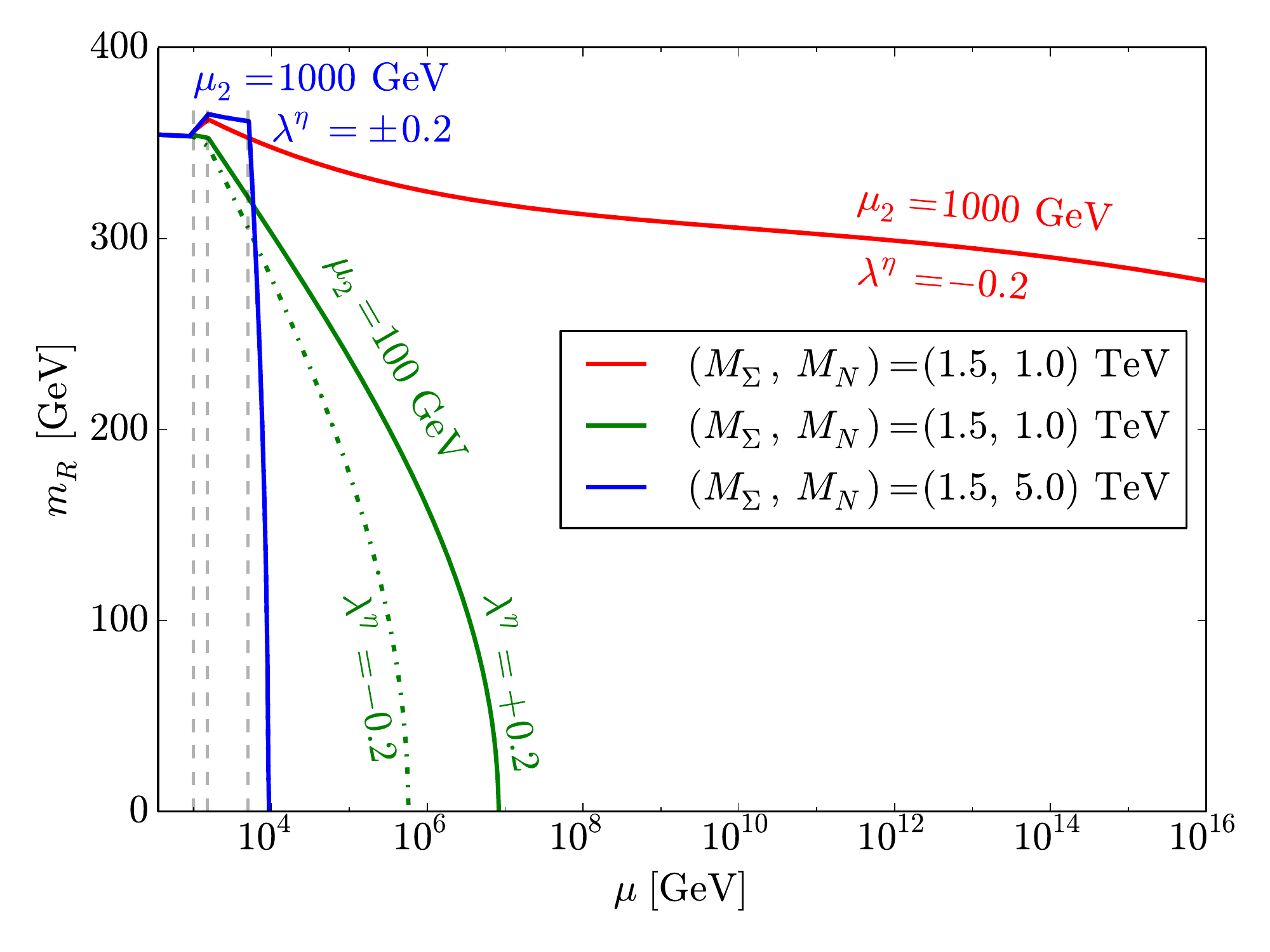}
%  \end{subfigure}
\caption{\label{fig:RunningMass}Running of the combinations of scalar quartic couplings relevant for the potential to be bounded from below (left panel) and of lightest inert scalar mass $m_R$ (right panel). Vertical dashed lines are particle thresholds.}
\end{figure}

First, we would like to direct the readers attention to FIG.~\ref{fig:RunningMass}, where the running of the conditions~\eqref{eq:boundedfrombelow} (left panel) and the lightest inert scalar mass parameter (right panel) is shown. The different colours in the right panel correspond to different values of fermion masses as indicated in the plot, where a scalar triplet mass parameter $m_\Omega^2 = - (900\,\mathrm{GeV})^2$ has been chosen. Here, a negative $m_\Omega^2$ is required by virtue of the tadpole equation~\eqref{eq:tad2}: Since we must have $v_\Omega \ll v_\phi$, either $\lambda_{1,2}^\Omega$ need to be very large, making the setting non-perturbative, or $m_\Omega^2$ and/or $\mu_1$ must be negative to solve the tadpole equation. However, applying the tadpole equations to the charged scalar mass matrix, we find that the physical charged Higgs mass $m_{H^\pm}^2 \sim \frac{\mu_1}{v_\Omega}$, and thus $\mu_1>0$ is required. Consequently, we need $m_\Omega^2<0$ to realise large triplet masses. In addition, we have verified that the conditions~\eqref{eq:boundedfrombelow} are never violated for the examples shown. As an illustration, the left panel of FIG.~\ref{fig:RunningMass} shows the running of the bounded-from-below conditions, see Eqs.~\eqref{eq:boundedfrombelow}, for one of the settings in the right panel (solid green line).

% {\color{red}
% \begin{itemize}
% \item{Description of the problem in terms of the numerical calculations}
% \item{The numerical settings taken in our study of the \z2 problem in the 
% parameter space: parameter ranges}
% \item{Results, plots and comments}
% \end{itemize}}

It can be concluded from FIG.~\ref{fig:RunningMass} that the situation is similar to the simplest scotogenic model in the sense that, once the heavy fermions become dynamic (i.e., above the renormalisation scale $\mu\ge M_{\Sigma / N}$), the RGEs of the inert mass $m_R$ contain large and negative terms that may eventually drive $m_R^2$ to negative values and induce $\mathbb{Z}_2$ breaking, cf.~the last two terms in Eq.~\eqref{eq:RGE_m_eta}:
\begin{equation}\label{eq:RGE_m_eta_short}
    \beta_{m_\eta^2} \sim -3 \lambda^\eta m_\Omega^2 +3 \mu_{2}^{2} +2 \Big(m_\eta^2 -2 |M_N|^2 \Big)\Big({Y_N  Y_N^*}\Big)
      +3 \Big( m_\eta^2 -2 |M_\Sigma|^2 \Big)\Big({Y_\Sigma  Y_\Sigma^*}\Big).
\end{equation}
Exactly that behaviour is the reflection of the parity problem in the singlet-triplet scotogenic model. However, there is a substantial difference with respect to the simplest scotogenic scenario, namely the presence of a scalar triplet field $\Omega$ which can counteract this effect. The interplay of fermion and scalar masses is manifest in the RGE~\eqref{eq:RGE_m_eta}, where in addition to the (generically negative) fermionic contributions, there are other contributions such as $\beta_{m_\eta^2} \sim -3 \lambda^\eta m_\Omega^2$. Depending on the sign of this contribution the breaking of $\mathbb{Z}_2$ can occur at higher scales or can be evaded all together. This behaviour can be clearly observed for the green curves in FIG.~\ref{fig:RunningMass}, but the effect is limited if $\lambda^\eta$ is restricted to magnitudes in the perturbative regime. More importantly, the dimensionful triple scalar couplings $\mu_{1,2}$ yield potentially large and \emph{positive} contributions to Eq.~\eqref{eq:RGE_m_eta_short}. The relevant term for the running of the inert scalar masses reads $\beta_{m_\eta^2} \sim + 3 \mu_2^2$. For a sufficiently large $\mu_2$, this contribution can outweigh that of the fermions $N$ and $\Sigma$, such that the scheme can remain consistent up to very high scales, as illustrated by the red curve in FIG.~\ref{fig:RunningMass}. Note that even though we have increased $\mu_2$ significantly, the effect on the physical mass is negligible. This is due to the fact that $\mu_2$ enters the relation for the physical masses~(\ref{eq:etaRmass}-\ref{eq:etaChargedmass}) multiplied by $v_\Omega$, which is forced to be very small. Finally, if the fermionic contributions dominate, as for the blue curve with $M_N=5\,\mathrm{TeV}$, the scalar contributions are practically irrelevant.
\begin{figure}[t]
  \centering
%  \begin{subfigure}{.45\textwidth}
%    \centering
    \includegraphics[width=0.49\textwidth]{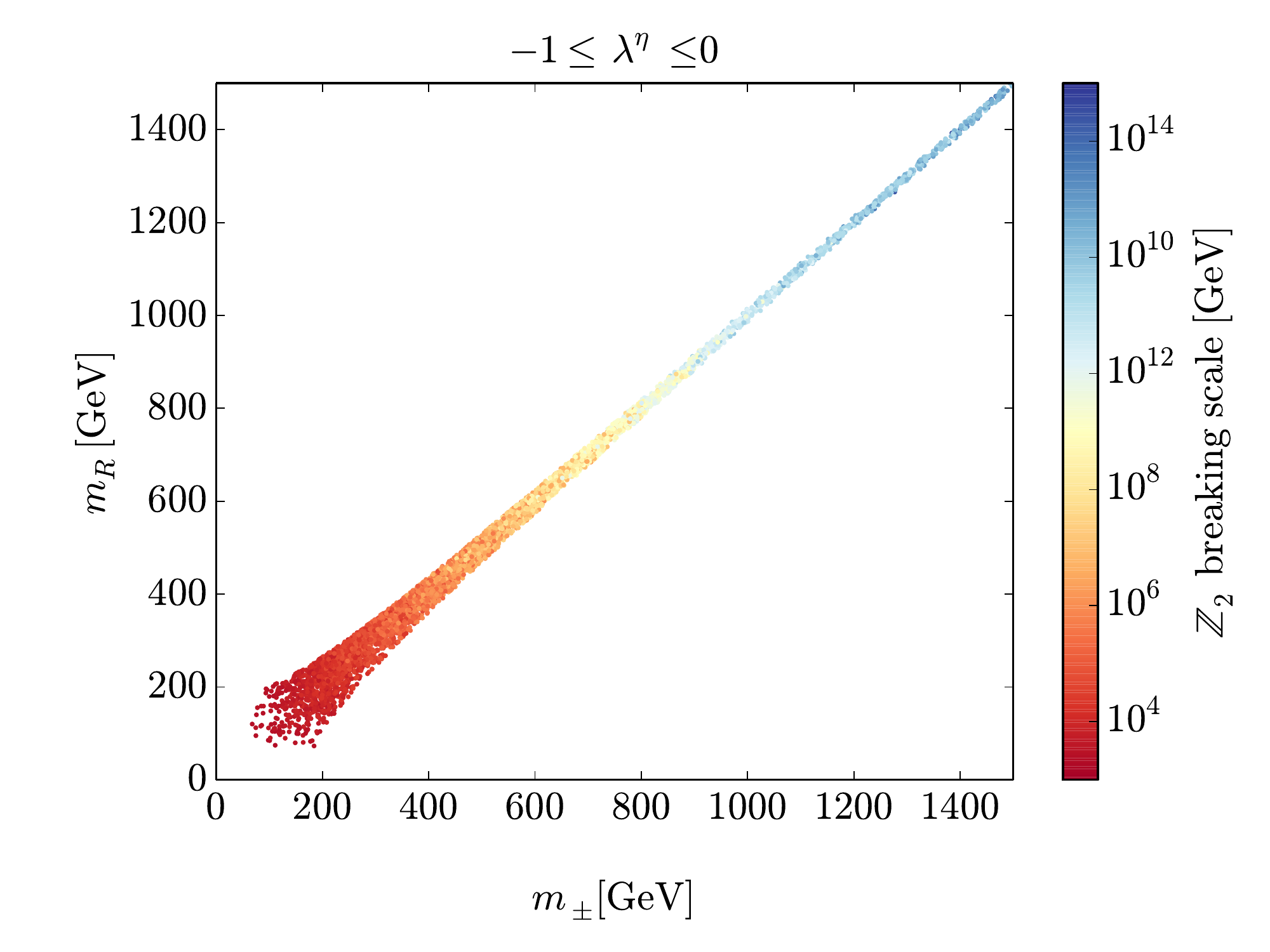}
%    \subcaption{(a)\label{fig:LambdaEtaNeg}$-0.1 \le \lambda^\eta \le 1$}
%  \end{subfigure}
%   
 \hfill
%   
%  \begin{subfigure}{.45\textwidth}
%    \centering
    \includegraphics[width=.49\textwidth]{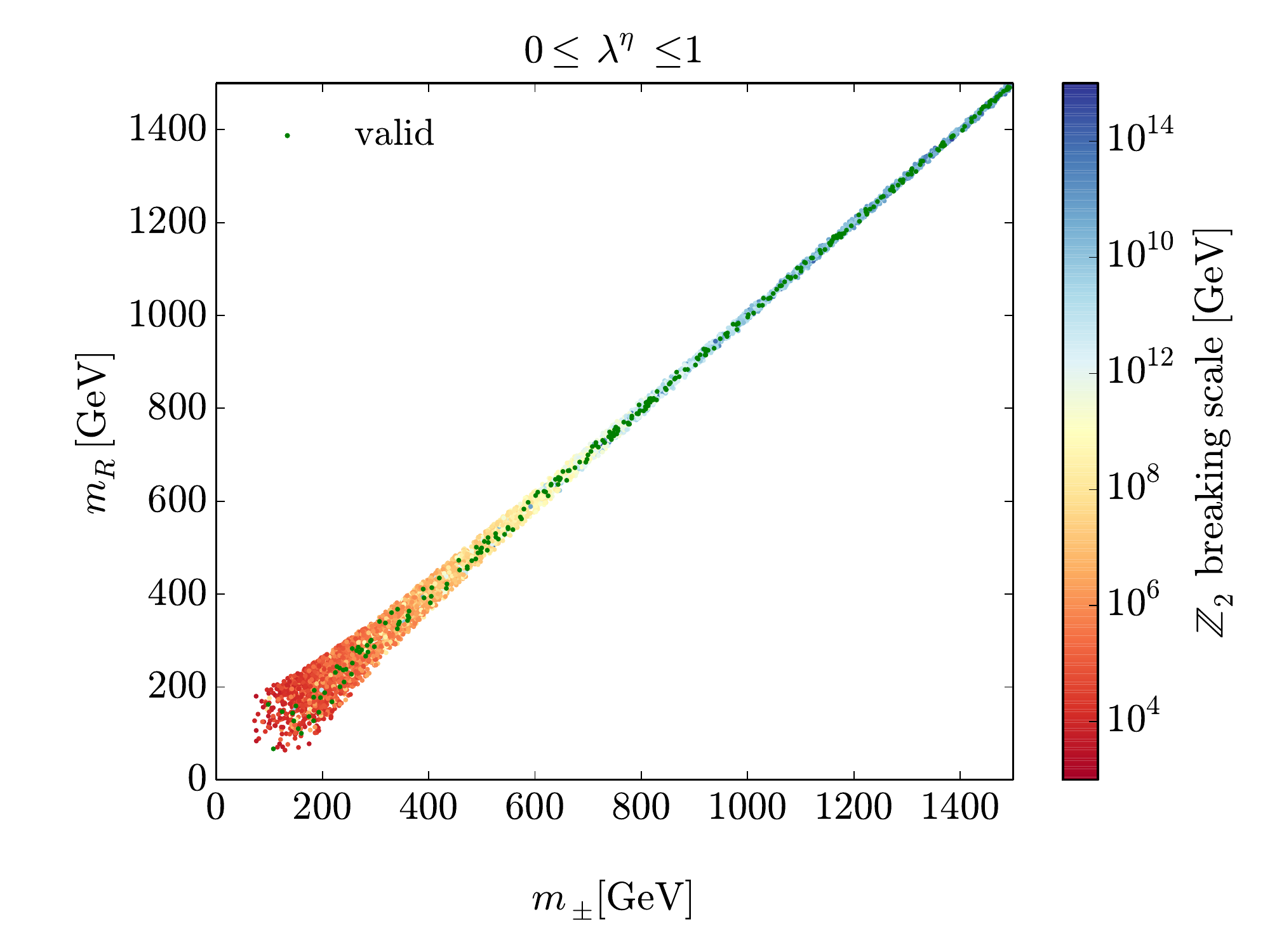}
  \caption{\label{fig:ParameterScanEta}Parameter scan of the model for different ranges of $\lambda^\eta$.}
\end{figure}

In order to better understand the impact of the running effects on the parameter space, we show in FIG.~\ref{fig:ParameterScanEta} a parameter scan of the model in the $m_{\eta^R}$-$m_{\eta^\pm}$ plane. To this end, we have chosen to fix the following parameters:
\begin{equation}
  M_\Sigma =1.5\,\textrm{TeV},\ M_N = 2.0\,\textrm{TeV},\ \lambda_2 = 0.2,\ \lambda_5 = 10^{-9},\ Y_\Omega =0.3,
\end{equation}
while $m_\phi$, $\lambda_1$, and $\mu_1$ are fixed by the tadpole equations for $v_\Omega = 0.5 \, \mathrm{GeV}$, and the requirement of finding a $125\,\mathrm{GeV}$ CP-even scalar in the spectrum, which is identified with the Higgs boson. The Yukawa couplings $Y_N$ and $Y_\Sigma$ are chosen according to an adapted Casas-Ibarra-parametrisation~\cite{Casas:2001sr} for one massless generation of neutrinos. The remaining parameters are varied in the following ranges generating a total of $50\,000$ points:
\begin{gather*}
    (100\,\mathrm{GeV})^2 \le m_\eta^2 \le (1500\,\mathrm{GeV})^2,\quad
    -(1500\,\mathrm{GeV})^2 \le m_\Omega^2 \le -(500\,\mathrm{GeV})^2,\\       
    -1 \le \lambda_3,\lambda_4,\lambda_1^\Omega \le 1,\quad
    0 \le \lambda_2^\Omega \le 1, \quad 0 \le \mu_2 \le 100\, \mathrm{GeV}.
\end{gather*}
The range of $\lambda^\eta$ has been chosen differently for the left and right panels of FIG.~\ref{fig:ParameterScanEta}, as given above each figure. We terminate the running at a scale $\Lambda = 10^{16}\,\mathrm{GeV}$ motivated by theories of grand unification. However, this is a merely practical choice and just as good as any other high scale, since no gauge coupling unification is required in this model. Any parameter point that runs up to this scale is considered valid and marked as a green point. Parameter combinations violating the bounded from below conditions~\eqref{eq:boundedfrombelow} or perturbativity are excluded from the plot. The remaining points indicate the breaking of $\mathbb{Z}_2$ and the corresponding scale at which the breaking occurs is displayed with a colour scale.

Quite generally, we see from FIG.~\ref{fig:ParameterScanEta} that the $\mathbb{Z}_2$ breaking scale rises with the inert masses, as expected. However, due to the large parameter space, the variation of the breaking scale for a given combination of masses is sizeable. Most notably, we see that if, $\lambda^\eta>0$, we are able to find many viable settings for almost all values of the masses $m_{\eta^R}$ and $m_{\eta^\pm}$. In contrast, restricting $\lambda^\eta$ to negative values \emph{no} viable setting is found. The reason for this is that the breaking scale of $\mathbb{Z}_2$ is now generally lowered by the scalar triplet contribution to the running of $m_\eta^2$, as highlighted in FIG.~\ref{fig:RunningMass}.

\begin{figure}[t]
  \centering
%  \begin{subfigure}{.45\textwidth}
%    \centering
    \includegraphics[width=0.49\textwidth]{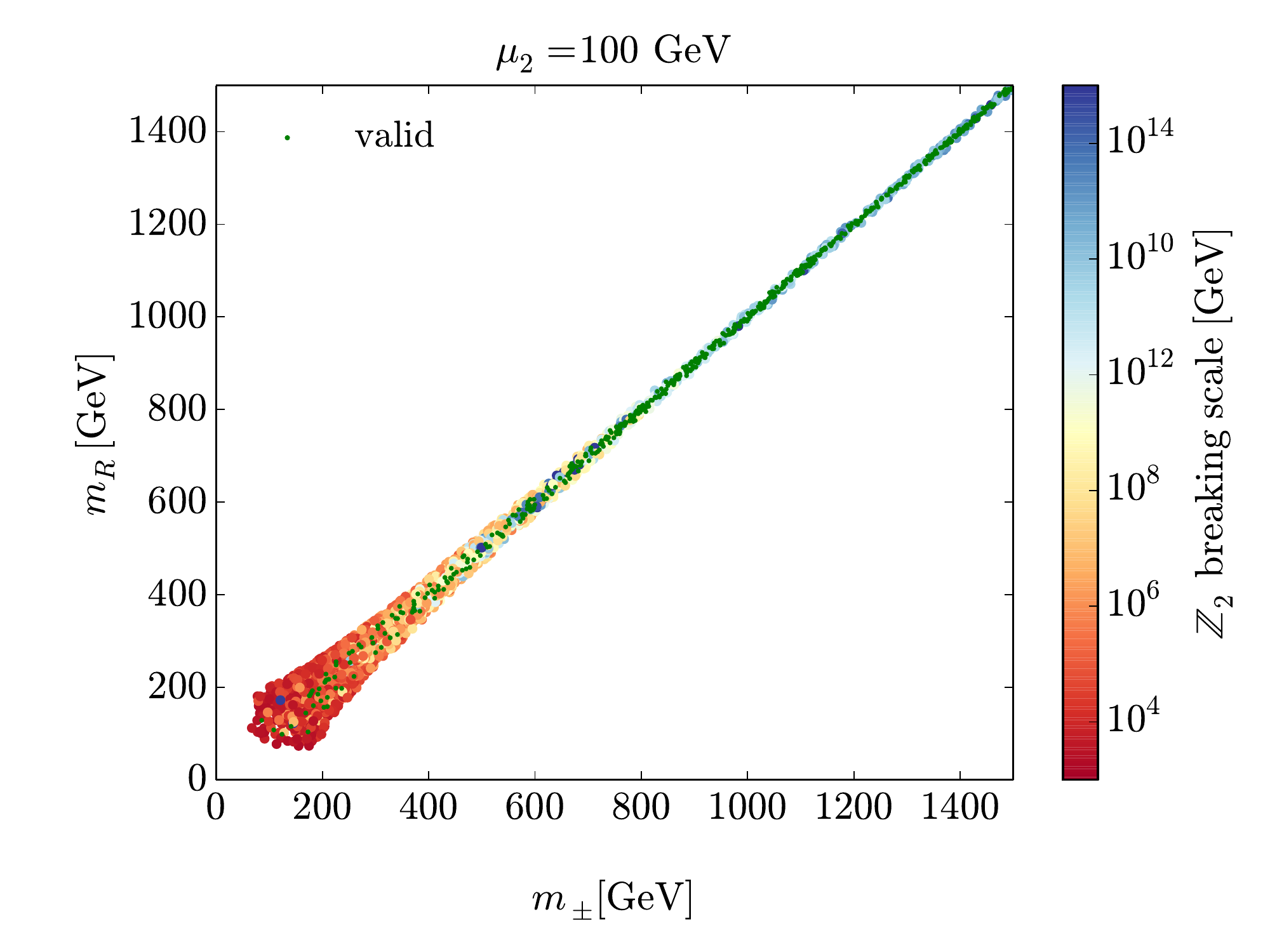}
%    \subcaption{(a)\label{fig:mu2_100}$\mu_2 = 100\,\mathrm{GeV}$}
%  \end{subfigure}
%   
 \hfill
%   
%  \begin{subfigure}{.45\textwidth}
%    \centering
    \includegraphics[width=.49\textwidth]{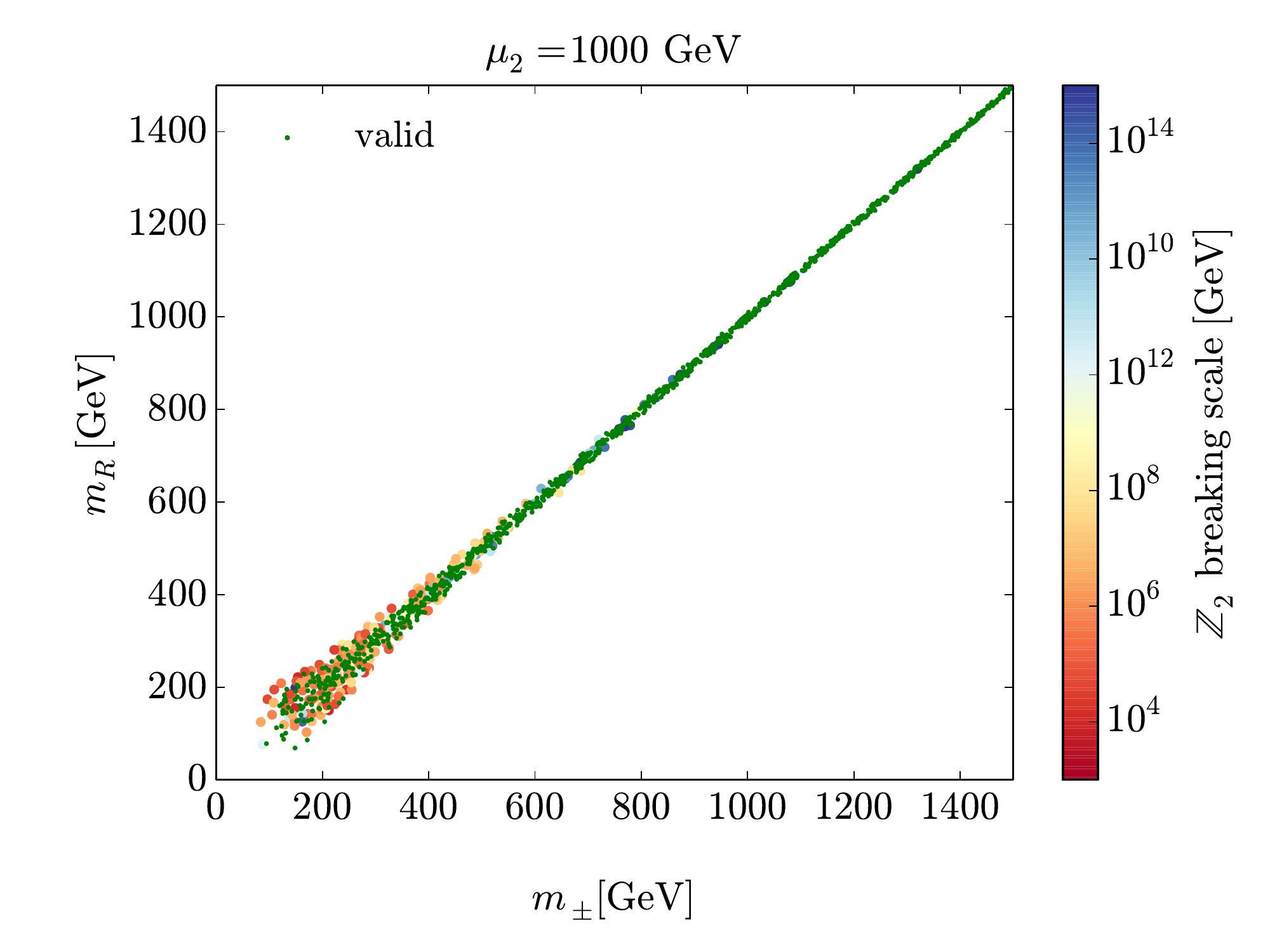}
%    \subcaption{(b)\label{fig:mu2_1000}$\mu_2 = 1000\,\mathrm{GeV}$}
%  \end{subfigure}
    \caption{\label{fig:ParameterScanMu}Parameter scan of the model
      for $\mu_2=100\,\mathrm{GeV}$ and $\mu_2=1\,\mathrm{TeV}$.}
\end{figure}

Similarly, glancing at FIG.~\ref{fig:ParameterScanMu} where we keep $\mu_2$ fixed and vary $-1 \le \lambda^\eta \le 1$ at the input scale, one observes that the impact of very large $\mu_2$ is as anticipated. For $\mu_2=100\,\mathrm{GeV}$ (left panel), $\mathbb{Z}_2$ breaking occurs for most of the points with inert scalar masses $\lesssim 500\,\mathrm{GeV}$. However, for $\mu_2 = 1\,\mathrm{TeV}$ most of the points turn out to be valid, even for such low scalar masses.\footnote{The choice $\mu_2 = 1\,\mathrm{TeV}$ is in fact quite natural, given that the RGE~\eqref{eq:RGE_mu_2} contains the fermion masses $M_{\Sigma/N}$.} Simultaneously, the overall scalar mass scale is unchanged due to $\mu_2$ entering the physical masses suppressed by the small triplet VEV.
 
In conclusion, the coupling $\lambda^\eta$ in combination with the mass scale of the scalar triplet, and the dimensionful scalar coupling $\mu_2$ may counteract the typical fermionic corrections to the inert scalar masses. Thus, they are the crucial ingredients that can naturally save the model from running into inconsistencies due to the breaking of the parity symmetry and provide a motivation for the presence of additional \emph{bosonic} degrees of freedom in scotogenic-type models.

\section{\label{sec:conclusions}Conclusions}

In this paper we have re-visited the scotogenic scenario, as it provides a common approach to the Dark Matter and neutrino mass generation problems, in which the same symmetry that stabilises Dark Matter also ensures the radiative seesaw origin of neutrino mass. We have carefully considered the behaviour of the required \z2 symmetry. In contrast to the simplest scenario, we have shown how the spontaneous breaking of \z2 can be naturally avoided in the singlet-triplet extension of the simplest model, up to fairly large energy scales, thanks to the presence of scalar triplets neutral under the \z2 which affect the evolution of the couplings in the scalar sector. The scenario offers good prospects for direct WIMP Dark Matter detection in nuclear recoil experiments, in ways quite analogous to supersymmetric Dark Matter stabilised by $R$-parity conservation.

\appendix

\section{Renormalisation Group Equations}
\label{app:RGEs}

The $\beta$ function of the parameter $c$, $\beta_c$, is defined by means of the renormalisation group equation
\begin{equation}
\frac{dc}{dt} = \beta_c = \sum_n \frac{1}{\left(16 \pi^2\right)^n} \, \beta_c^{(n)} \, ,
\end{equation}
where $t = \log \mu$, $\mu$ being the energy scale, and $\beta_c^{(n)}$ is the n-loop $\beta$ function. In this paper, we used {\tt SARAH}~\cite{Staub:2013tta,Staub:2015kfa} to compute the $\beta$ functions of all parameters in $R_\xi$ gauge at the 1-loop level. We summarise our results here. Notice that we drop the superindex $^{(1)}$ for the sake of clarity.

\input{rges}

\section*{Acknowledgements}

AM acknowledges partial support by the European Union FP7 ITN INVISIBLES (Marie Curie Actions, PITN-GA-2011-289442) and by the Micron Technology Foundation, Inc. This work is supported by the Spanish grants FPA2014-58183-P, Multidark CSD2009-00064, SEV-2014-0398 (MINECO) and PROMETEOII/2014/084 (Generalitat Valenciana). MP acknowledges support from the IMPRS-PTFS. NR was funded by becas de postdoctorado en el extranjero Conicyt/Becas Chile 74150028. AV acknowledges financial support from the ``Juan de la Cierva'' program (27-13-463B- 731) funded by the Spanish MINECO. 

\bibliography{merged_Valle.bib,newrefs.bib}

\end{document}

%% file: rges.tex
\subsection{Gauge Couplings}
{\allowdisplaybreaks  \begin{align} 
\beta_{g_1} & =  
\frac{21}{5} g_{1}^{3} \\ 
\beta_{g_2} & =  
-\frac{4}{3} g_{2}^{3} \\ 
\beta_{g_3} & =  
-7 g_{3}^{3}
\end{align}} 
\subsection{Quartic scalar couplings}
{\allowdisplaybreaks  \begin{align} 
\beta_{\lambda_1} & =  
+\frac{27}{100} g_{1}^{4} +\frac{9}{10} g_{1}^{2} g_{2}^{2} +\frac{9}{4} g_{2}^{4} -\frac{9}{5} g_{1}^{2} \lambda_1 -9 g_{2}^{2} \lambda_1 +12 \lambda_{1}^{2} +4 \lambda_{3}^{2} +4 \lambda_3 \lambda_4 +2 \lambda_{4}^{2} +2 \lambda_{5}^{2} + 3 \left( \lambda_1^\Omega \right)^2 \nonumber \\ 
 &+12 \lambda_1 \mbox{Tr}\Big({Y_d  Y_{d}^{\dagger}}\Big) +4 \lambda_1 \mbox{Tr}\Big({Y_e  Y_{e}^{\dagger}}\Big) +12 \lambda_1 \mbox{Tr}\Big({Y_u  Y_{u}^{\dagger}}\Big) -12 \mbox{Tr}\Big({Y_d  Y_{d}^{\dagger}  Y_d  Y_{d}^{\dagger}}\Big) -4 \mbox{Tr}\Big({Y_e  Y_{e}^{\dagger}  Y_e  Y_{e}^{\dagger}}\Big) \nonumber \\ 
 &-12 \mbox{Tr}\Big({Y_u  Y_{u}^{\dagger}  Y_u  Y_{u}^{\dagger}}\Big) \\ 
\beta_{\lambda_2} & =  
+\frac{27}{100} g_{1}^{4} +\frac{9}{10} g_{1}^{2} g_{2}^{2} +\frac{9}{4} g_{2}^{4} -\frac{9}{5} g_{1}^{2} \lambda_2 -9 g_{2}^{2} \lambda_2 +12 \lambda_{2}^{2} +4 \lambda_{3}^{2} +4 \lambda_3 \lambda_4 +2 \lambda_{4}^{2} +2 \lambda_{5}^{2} + 3 \left( \lambda^\eta \right)^2 \nonumber \\ 
 &+4 \lambda_2 \Big({Y_N  Y_N^*}\Big) -4 \Big({Y_N  Y_N^*}\Big)^{2} -4 \Big({Y_N  Y_\Sigma^*}\Big) \Big({Y_\Sigma  Y_N^*}\Big) +6 \lambda_2 \Big({Y_\Sigma  Y_\Sigma^*}\Big) -5 \Big({Y_\Sigma  Y_\Sigma^*}\Big)^{2} \\ 
\beta_{\lambda_3} & =  
+\frac{27}{100} g_{1}^{4} -\frac{9}{10} g_{1}^{2} g_{2}^{2} +\frac{9}{4} g_{2}^{4} -\frac{9}{5} g_{1}^{2} \lambda_3 -9 g_{2}^{2} \lambda_3 +6 \lambda_1 \lambda_3 +6 \lambda_2 \lambda_3 +4 \lambda_{3}^{2} +2 \lambda_1 \lambda_4 +2 \lambda_2 \lambda_4 +2 \lambda_{4}^{2} \nonumber \\ 
 &+2 \lambda_{5}^{2} + 3 \lambda^\Omega_1 \lambda^\eta +2 \lambda_3 \Big({Y_N  Y_N^*}\Big) -4 \Big({Y_N  {Y_{e}^{\dagger}  Y_e  Y_N^*}}\Big) +3 \lambda_3 \Big({Y_\Sigma  Y_\Sigma^*}\Big) -2 \Big({Y_\Sigma  {Y_{e}^{\dagger}  Y_e  Y_\Sigma^*}}\Big) \nonumber \\ 
 &+6 \lambda_3 \mbox{Tr}\Big({Y_d  Y_{d}^{\dagger}}\Big) +2 \lambda_3 \mbox{Tr}\Big({Y_e  Y_{e}^{\dagger}}\Big) +6 \lambda_3 \mbox{Tr}\Big({Y_u  Y_{u}^{\dagger}}\Big) \\ 
\beta_{\lambda_4} & =  
+\frac{9}{5} g_{1}^{2} g_{2}^{2} -\frac{9}{5} g_{1}^{2} \lambda_4 -9 g_{2}^{2} \lambda_4 +2 \lambda_1 \lambda_4 +2 \lambda_2 \lambda_4 +8 \lambda_3 \lambda_4 +4 \lambda_{4}^{2} +8 \lambda_{5}^{2} +2 \lambda_4 \Big({Y_N  Y_N^*}\Big) \nonumber \\ 
 &+4 \Big({Y_N  {Y_{e}^{\dagger}  Y_e  Y_N^*}}\Big) +3 \lambda_4 \Big({Y_\Sigma  Y_\Sigma^*}\Big) -2 \Big({Y_\Sigma  {Y_{e}^{\dagger}  Y_e  Y_\Sigma^*}}\Big) +6 \lambda_4 \mbox{Tr}\Big({Y_d  Y_{d}^{\dagger}}\Big) +2 \lambda_4 \mbox{Tr}\Big({Y_e  Y_{e}^{\dagger}}\Big) \nonumber \\ 
 &+6 \lambda_4 \mbox{Tr}\Big({Y_u  Y_{u}^{\dagger}}\Big) \\ 
\beta_{\lambda_5} & =  
-\frac{9}{5} g_{1}^{2} \lambda_5 -9 g_{2}^{2} \lambda_5 +2 \lambda_1 \lambda_5 +2 \lambda_2 \lambda_5 +8 \lambda_3 \lambda_5 +12 \lambda_4 \lambda_5 +2 \lambda_5 \Big({Y_N  Y_N^*}\Big) +3 \lambda_5 \Big({Y_\Sigma  Y_\Sigma^*}\Big) \nonumber \\ 
 &+6 \lambda_5 \mbox{Tr}\Big({Y_d  Y_{d}^{\dagger}}\Big) +2 \lambda_5 \mbox{Tr}\Big({Y_e  Y_{e}^{\dagger}}\Big) +6 \lambda_5 \mbox{Tr}\Big({Y_u  Y_{u}^{\dagger}}\Big) \label{eq:RGE_lambda_5} \\ 
\beta_{\lambda^\Omega_1} & =  
+3 g_{2}^{4} -\frac{9}{10} g_{1}^{2} \lambda^\Omega_1 -\frac{33}{2} g_{2}^{2} \lambda^\Omega_1 +6 \lambda_1 \lambda^\Omega_1 +4 \left( \lambda_1^\Omega \right)^2 + 10 \lambda^\Omega_1 \lambda^\Omega_2 +4 \lambda_3 \lambda^\eta +2 \lambda_4 \lambda^\eta +4 \lambda^\Omega_1 |Y_\Omega|^2 \nonumber \\ 
 &+6 \lambda^\Omega_1 \mbox{Tr}\Big({Y_d  Y_{d}^{\dagger}}\Big) +2 \lambda^\Omega_1 \mbox{Tr}\Big({Y_e  Y_{e}^{\dagger}}\Big) +6 \lambda^\Omega_1 \mbox{Tr}\Big({Y_u  Y_{u}^{\dagger}}\Big) \\ 
\beta_{\lambda^\Omega_2} & =  
-24 g_{2}^{2} \lambda^\Omega_2 + 22 \left( \lambda_2^\Omega \right)^2  + 6 g_{2}^{4}  + 8 \lambda^\Omega_2 |Y_\Omega|^2  -8 |Y_\Omega|^4  + 2 \left( \lambda_1^\Omega \right)^2 + 2 \left( \lambda^\eta \right)^2\\  
\beta_{\lambda^\eta} & =  
+3 g_{2}^{4} +4 \lambda_3 \lambda^\Omega_1 +2 \lambda_4 \lambda^\Omega_1 -\frac{9}{10} g_{1}^{2} \lambda^\eta -\frac{33}{2} g_{2}^{2} \lambda^\eta +6 \lambda_2 \lambda^\eta +10 \lambda^\Omega_2 \lambda^\eta +4 \left( \lambda^\eta \right)^2 +2 \lambda^\eta \Big({Y_N  Y_N^*}\Big) \nonumber \\ 
 &+4 |Y_\Omega|^2 \Big(\lambda^\eta -2 \Big({Y_N  Y_N^*}\Big)  - \Big({Y_\Sigma  Y_\Sigma^*}\Big) \Big)+3 \lambda^\eta \Big({Y_\Sigma  Y_\Sigma^*}\Big)
\end{align}} 
\subsection{Yukawa Couplings}
{\allowdisplaybreaks  \begin{align} 
\beta_{Y_u^{\alpha \beta}} & =  
\frac{3}{2} \Big( {Y_u  Y_{u}^{\dagger}  Y_u} - {Y_u  Y_{d}^{\dagger}  Y_d}\Big)^{\alpha \beta} \nonumber \\ 
 &+ \Big(3 \mbox{Tr}\Big({Y_d  Y_{d}^{\dagger}}\Big)  + 3 \mbox{Tr}\Big({Y_u  Y_{u}^{\dagger}}\Big) + \mbox{Tr}\Big({Y_e  Y_{e}^{\dagger}}\Big) - \frac{17}{20} g_{1}^{2}  -\frac{9}{4} g_{2}^{2}  -8 g_{3}^{2} \Big) \, Y_u^{\alpha \beta} \\ 
\beta_{Y_d^{\alpha \beta}} & =  
\frac{3}{2} \Big( {Y_d  Y_{d}^{\dagger}  Y_d} - {Y_d  Y_{u}^{\dagger}  Y_u}  \Big)^{\alpha \beta} \nonumber \\ 
 &+ \Big(3 \mbox{Tr}\Big({Y_d  Y_{d}^{\dagger}}\Big)  + 3 \mbox{Tr}\Big({Y_u  Y_{u}^{\dagger}}\Big) + \mbox{Tr}\Big({Y_e  Y_{e}^{\dagger}}\Big) - \frac{1}{4} g_{1}^{2} - \frac{9}{4} g_{2}^{2} -8 g_{3}^{2} \Big) \, Y_d^{\alpha \beta} \\ 
\beta_{Y_e^{\alpha \beta}} & =  
\frac{3}{2} {Y_e  Y_{e}^{\dagger}  Y_e} + \frac{1}{2} \Big({Y_e  Y_N^*}\Big)^\alpha Y_N^\beta + \frac{3}{4} \Big({Y_e  Y_\Sigma^*}\Big)^\alpha Y_\Sigma^\beta \nonumber \\ 
&+ \Big(3 \mbox{Tr}\Big({Y_d  Y_{d}^{\dagger}}\Big)  + 3 \mbox{Tr}\Big({Y_u  Y_{u}^{\dagger}}\Big)  + \mbox{Tr}\Big({Y_e  Y_{e}^{\dagger}}\Big)  -\frac{9}{4} g_{1}^{2}  -\frac{9}{4} g_{2}^{2} \Big) \, Y_e^{\alpha \beta} \\
\beta_{Y_N^\alpha} & =  
\Big(\frac{3}{2} |Y_\Omega|^2  + \frac{3}{2} \Big({Y_\Sigma  Y_\Sigma^*}\Big)  + \frac{5}{2} \Big({Y_N  Y_N^*}\Big) -\frac{9}{20} g_{1}^{2} -\frac{9}{4} g_{2}^{2} \Big) \, Y_N^\alpha  + \frac{1}{2} \Big({Y_{e}^{T}  Y_e^*  Y_N}\Big)^\alpha  + \frac{3}{4} \Big({Y_N  Y_\Sigma^*}\Big) Y_\Sigma^\alpha \label{eq:RGE_Y_N}\\ 
\beta_{Y_\Sigma^\alpha} & =  
\Big(\frac{1}{2} |Y_\Omega|^2  + \Big({Y_N  Y_N^*}\Big) + \frac{11}{4} \Big({Y_\Sigma  Y_\Sigma^*}\Big)  -\frac{9}{20} g_{1}^{2} -\frac{33}{4} g_{2}^{2} \Big) \, Y_\Sigma^\alpha +\frac{1}{2} \Big({Y_{e}^{T}  Y_e^*  Y_\Sigma}\Big)^\alpha  + \frac{1}{2} \Big({Y_\Sigma  Y_N^*}\Big) Y_N^\alpha  \label{eq:RGE_Y_Sigma}\\
\beta_{Y_\Omega} & =  
\Big(6 |Y_\Omega|^2  + \Big({Y_N  Y_N^*}\Big)  + \frac{1}{2} \Big({Y_\Sigma  Y_\Sigma^*}\Big) -6 g_{2}^{2} \Big) \, Y_\Omega  \label{eq:RGE_Y_Omega}
\end{align}} 
\subsection{Fermion Mass Terms}
{\allowdisplaybreaks  \begin{align} 
\beta_{M_N} & =  
2 M_N \Big({Y_N  Y_N^*}\Big)  + 3 M_N |Y_\Omega|^2  + 6 Y_{\Omega}^{2} M_\Sigma^* \label{eq:RGE_M_N}\\ 
\beta_{M_\Sigma} & =  
2 Y_{\Omega}^{2} M_N^*  + M_\Sigma \Big(-12 g_{2}^{2}  + |Y_\Omega|^2 + \Big({Y_\Sigma  Y_\Sigma^*}\Big)\Big) \label{eq:RGE_M_Sigma}
\end{align}} 
\subsection{Trilinear Scalar couplings}
{\allowdisplaybreaks  \begin{align} 
\beta_{\mu_1} & =  
-\frac{9}{10} g_{1}^{2} \mu_1 -\frac{21}{2} g_{2}^{2} \mu_1 +2 \lambda_1 \mu_1 +4 \lambda^\Omega_1 \mu_1 +2 \lambda_4 \mu_2 +2 \mu_1 |Y_\Omega|^2 +6 \mu_1 \mbox{Tr}\Big({Y_d  Y_{d}^{\dagger}}\Big) +2 \mu_1 \mbox{Tr}\Big({Y_e  Y_{e}^{\dagger}}\Big) \nonumber \\ 
 &+6 \mu_1 \mbox{Tr}\Big({Y_u  Y_{u}^{\dagger}}\Big) \\ 
\beta_{\mu_2} & =  
+2 \lambda_4 \mu_1 -\frac{9}{10} g_{1}^{2} \mu_2 -\frac{21}{2} g_{2}^{2} \mu_2 +2 \lambda_2 \mu_2 +4 \lambda^\eta \mu_2 +2 \mu_2 \Big({Y_N  Y_N^*}\Big) +4 Y_\Omega M_N^* \Big({Y_N  Y_\Sigma^*}\Big) +4 Y_\Omega M_\Sigma^* \Big({Y_\Sigma  Y_N^*}\Big) \nonumber \\ 
 &+2 Y_\Omega^* \Big(2 M_N \Big({Y_\Sigma  Y_N^*}\Big)  + 2 M_\Sigma \Big({Y_N  Y_\Sigma^*}\Big)  + \mu_2 Y_\Omega \Big)+3 \mu_2 \Big({Y_\Sigma  Y_\Sigma^*}\Big) \label{eq:RGE_mu_2}
\end{align}} 
\subsection{Scalar Mass Terms}
{\allowdisplaybreaks  \begin{align} 
\beta_{m_\phi^2} & =  
-4 \lambda_3 m_\eta^2 -2 \lambda_4 m_\eta^2 -\frac{9}{10} g_{1}^{2} m_\phi^2 -\frac{9}{2} g_{2}^{2} m_\phi^2 +6 \lambda_1 m_\phi^2 + 3 \lambda^\Omega_1 m_\Omega^2 -3 \mu_{1}^{2} +6 m_\phi^2 \mbox{Tr}\Big({Y_d  Y_{d}^{\dagger}}\Big) \nonumber \\ 
 &+2 m_\phi^2 \mbox{Tr}\Big({Y_e  Y_{e}^{\dagger}}\Big) +6 m_\phi^2 \mbox{Tr}\Big({Y_u  Y_{u}^{\dagger}}\Big) \\ 
\beta_{m_\eta^2} & =  
-\frac{9}{10} g_{1}^{2} m_\eta^2 -\frac{9}{2} g_{2}^{2} m_\eta^2 +6 \lambda_2 m_\eta^2 -4 \lambda_3 m_\phi^2 -2 \lambda_4 m_\phi^2 - 3 \lambda^\eta m_\Omega^2 +3 \mu_{2}^{2} +2 \Big(m_\eta^2 -2 |M_N|^2 \Big)\Big({Y_N  Y_N^*}\Big) \nonumber \\ 
 &+3 \Big( m_\eta^2 -2 |M_\Sigma|^2 \Big)\Big({Y_\Sigma  Y_\Sigma^*}\Big) \label{eq:RGE_m_eta} \\ 
\beta_{m_\Omega^2} & =  
-2 \Big(2 \lambda^\eta m_\eta^2 -2 \lambda^\Omega_1 m_\phi^2 +6 g_{2}^{2} m_\Omega^2 - 5 \lambda^\Omega_2 m_\Omega^2 +\mu_{1}^{2}+\mu_{2}^{2}-2 |Y_\Omega|^2 \Big(2 M_\Sigma M_\Sigma^*  + m_\Omega^2\Big)-2 M_N M_\Sigma Y_{\Omega}^{*,2} \nonumber \\ 
 &-2 Y_\Omega M_N^* \Big(2 M_N Y_\Omega^*  + Y_\Omega M_\Sigma^* \Big)\Big)
\end{align}}